\documentclass[manuscript]{aastex62}

\submitjournal{AAS Research Notes}
\shorttitle{A Possible Cold Stellar Stream in the Ultra-Diffuse Galaxy NGC1052-DF2}
\shortauthors{Abraham et al.}

\begin{document}

\title{{\em The Maybe Stream}: A Possible Cold Stellar Stream in the Ultra-Diffuse Galaxy NGC1052-DF2}
\correspondingauthor{Roberto Abraham}
\email{abraham@astro.utoronto.ca}

\author{Roberto Abraham}
\affil{Department of Astronomy \& Astrophysics, University of Toronto, 50 St. George Street, Toronto, ON M5S 3H4, Canada}
\affil{Dunlap Institute for Astronomy \& Astrophysics, 50 St. George Street, Toronto, ON M5S 3H4, Canada}

\author{Shany Danieli}
\affiliation{Astronomy Department, Yale University, New Haven, CT 06511, USA}

\author{Pieter van Dokkum}
\affiliation{Astronomy Department, Yale University, New Haven, CT 06511, USA}

\author{Charlie Conroy}
\affiliation{Harvard-Smithsonian Center for Astrophysics, 60 Garden Street, Cambridge, MA, USA}

\author{J. M. Diederik Kruijssen}
\affil{Astronomisches Rechen-Institut, Zentrum f\"ur Astronomie der Universit\"at Heidelberg, M\"onchhofstrasse 12-14, D-69120 Heidelberg, Germany}

\author{Yotam Cohen}
\affiliation{Astronomy Department, Yale University, New Haven, CT 06511, USA}

\author{Allison Merritt}
\affiliation{Max-Planck-Institut f\"ur Astronomie, K\"onigstuhl 17, D-69117 Heidelberg, Germany}

\author{Jielai Zhang}
\affil{Department of Astronomy \& Astrophysics, University of Toronto, 50 St. George Street, Toronto, ON M5S 3H4, Canada}
\affil{Dunlap Institute for Astronomy \& Astrophysics, 50 St. George Street, Toronto, ON M5S 3H4, Canada}

\author{Deborah Lokhorst}
\affil{Department of Astronomy \& Astrophysics, University of Toronto, 50 St. George Street, Toronto, ON M5S 3H4, Canada}
\affil{Dunlap Institute for Astronomy \& Astrophysics, 50 St. George Street, Toronto, ON M5S 3H4, Canada}

\author{Lamiya Mowla}
\affiliation{Astronomy Department, Yale University, New Haven, CT 06511, USA}

\author{Jean Brodie}
\affiliation{University of California Observatories, 1156 High Street, Santa Cruz, CA 95064, USA}

\author{Aaron J. Romanowsky}
\affiliation{Department of Physics and Astronomy, San Jos\'e State University, San Jose, CA 95192, USA}

\author{Steven Janssens}
\affil{Department of Astronomy \& Astrophysics, University of Toronto, 50 St. George Street, Toronto, ON M5S 3H4, Canada}

%\begin{abstract}
%We report tentative evidence for a cold stellar stream in the
%ultra-diffuse galaxy NGC1052-DF2. 
%If confirmed, this stream (which we refer to as `The Maybe Stream') would be the first cold (width $\lesssim
%10$\,pc) stream detected outside
%of the Local Group. The candidate stream is very narrow and has an
%unusual and highly curved shape.
%\end{abstract}

\section{A Cold Stellar Stream in an Ultra-Diffuse Galaxy?} \label{sec:intro}

The nearby ultra-diffuse galaxy \citep[UDG;][]{vandokkum15}
NGC1052-DF2 has recently been the subject of two papers \citep{vandokkum18a,vandokkum18b}
 which focus on a lack of dark matter in this object (a subject of intense debate\footnote{Much of the discussion occurred on social media, but see \citet{martin18} for a critique, and 
\url{https://www.pietervandokkum.com/ngc1052-df2} for a response.}),
and on the properties of its globular cluster population. In this
Research Note we draw attention to another aspect of this 
galaxy: it contains a very thin `S-shaped' structure made up
of resolved stars (Figure 1). In this note we point out the
possibility that this structure is a cold stellar stream, which we refer
to as `The Maybe Stream'.

\begin{figure}[hbt]
\centering
\includegraphics[width=7.2in]{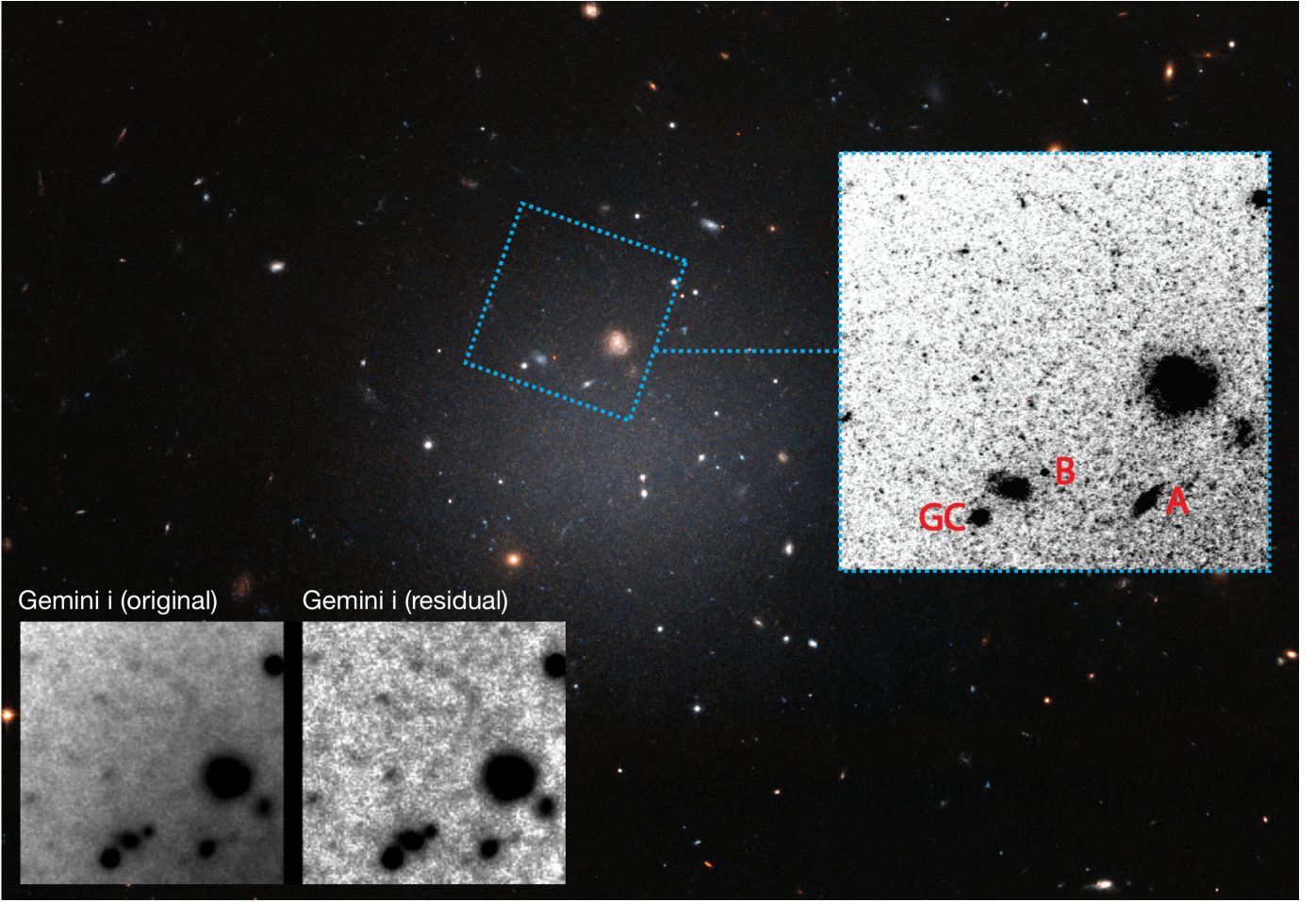}
\caption{\label{fig:stream}An image of NGC1052-DF2 constructed by
  combining the F814W and F606W data to represent
  `true' colours. The right-hand inset shows a portion of F814W image,
  centred on the candidate `S-shaped' (actually, $\wr$-shaped)
  cold stellar stream (`The Maybe Stream'). The candidate stream is clearly visible in both filters
  and portions of it are resolved into stars. The feature is also visible in a Gemini $i$ band image,
particularly after subtracting a model for the smooth light (bottom left).
Objects labeled `A' and  `B' 
 are candidates for the stream nucleus. The object marked `GC' is a globular
cluster (see \nocite{vandokkum18b} van Dokkum et al. 2018b). 
}
\end{figure}

The most distant cold stellar stream known is the Pisces/Triangulum
stream \citep{bonaca12,martin13}
at a distance of only
${\sim}35$ kpc. NGC1052-DF2 lies at a distance of ${\sim}20$ Mpc
\citep[see][]{vandokkum18a}
and is thus about 500$\times$ further away. A typical Milky
Way cold stream is ${\sim}10$ pc wide, which at a distance of ${\sim}20$
Mpc corresponds to a width of about 0.1 arcsec. This width is consistent
with that of the observed structure.

%If our interpretation of The Maybe Stream is
%correct, a stellar stream in NGC1052-DF2 is detectable at such a large
%distance mainly because very thin resolved structures that are ordinarily
%invisible when seen in projection against conventional galaxies remain
%visible when seen against the low surface brightness of a UDG
%envelope.

Our interpretation of this structure as a cold stream is tentative.  Figure 1 is based on two orbit (one orbit per
filter) $F606W$ and $F814W$ Hubble Space Telescope ({\it HST}) imaging. The galaxy is marginally-resolved, and the putative stream is composed of many strong positive surface brightness fluctuations superposed on two opposing thin `arcs' of background light.
There is no obvious difference between the color of the candidate stream and 
that of the rest of the galaxy.

Three candidates for the nucleus of the stream are marked as `GC', `A',
and `B' on Figure 1.
Candidate `A' is  likely a background galaxy.
Candidate `B' is very intriguing.
It lies directly on the lower arc
of the candidate stream, and is one of the reddest objects on the entire
ACS image, with $(F606W - F814W) \sim 3.5$ mag.
It has $F814W \sim 23.5$ mag, and is not resolved.
%It could be the core of the putative stream, although in that case it
%is hard to understand why it is so red. 
While very red, it is almost certainly
not a TP-AGB star, as it is about three magnitudes brighter than the
tip of the RGB.
Object `GC' is the globular cluster GC-85 \citep[see][]{vandokkum18b}.
This globular cluster
is elongated in the same direction as the stream, but it does not appear to
be connected: the stream appears to curve to the South (toward object `B').
%On the other hand, this is the only one of the three candidates
%that has a stream-producing counterpart 
%in the Milky Way, and it could be that the turn in the stream is not real.
Finally, we note that the stream may be an ``orphan'' \citep[see][]{belokurov07}, where the progenitor object is either completely disrupted or
presently far removed from the densest region.

\section{Topology of the candidate stream} \label{sec:intro}

Cold streams are presumed to originate from disrupting dwarf galaxies
or globular clusters, and
if the candidate stream is real its abrupt curves would be very
remarkable.
The shapes of streams are sensitive to the
strength and shape of the gravitational potential \citep{helmi04,johnston05,lux12}, as well as to its substructure
\citep{carlberg09,yoon11,carlberg17}.
%Individual streams provide
%constraints on the potential of the galaxy causing the
%disruption, and the complexity of the streams increases in
%lock-step with the complexity of the substructure.
%Complicated streams
%that at least superficially resemble the structure seen in NGC1052-DF2
%are sometimes seen in simulations when the tidal field
%that the disrupting object experiences changes during its orbit
%(Carlberg 2017).  
%It would be
%interesting to see if numerical modelling is able to produce a complex
%stream topology in the absence of a dominant dark matter halo.
Once a stream is set up in a galaxy such as NGC1052-DF2, additional sharp turns might be the product of few-body dynamics, because the unusual properties of the galaxy make it more probable that a stream would overlap with the spheres of influence (SOI) of the individual globular clusters in the system. In the Milky Way, a typical globular cluster  has a mass of $2\times10^5~M_\odot$ and orbits at 5 kpc \citep{harris96}, with an enclosed Galactic mass of $5\times10^{10}~M_\odot$, resulting in an SOI radius of 33 pc. In NGC1052-DF2, the corresponding values are $8\times10^5~M_\odot$, with 2.5 kpc and $1\times10^8~M_\odot$ \citep{vandokkum18a}, resulting in clusters with an enormous SOI radius of 360 pc, about ten times larger than for clusters in the Milky Way. While strong accelerations are not expected at large distances within the SOI, these numbers make the point that strongly bent streams might be more commonly seen in some ultra-diffuse galaxies than in more conventional galaxies.

%We conclude this note by stressing that 
The currently-available data
are not deep enough to confirm the existence of the stream beyond
reasonable doubt.  The eye is very good at detecting patterns, and 
%we are aware that 
random clumps of giant stars can line up to form
phantom features. 
%Still, the structure seems remarkable enough to be
%worth noting, and 
Deeper data obtained with {\it HST} or {\it JWST} will be able
to either confirm or refute the reality of The Maybe Stream.

\end{document}